\documentclass[traditabstract]{aa} 
\usepackage{graphicx}
\usepackage{txfonts}
\usepackage{natbib,twoopt}
\usepackage{textcomp}
\usepackage{longtable}
\usepackage{booktabs}
\usepackage{lscape}
\usepackage{xspace}
\usepackage{tabularx} 
\begin{document}
\title{Modelling photometric reverberation data -- a disk-like broad-line region and a potentially larger black hole mass for 3C120.}
\titlerunning{Modelling photometric reverberation data of 3C120}

\author{
  F. Pozo Nu\~nez 
  \inst{1} 
  \and
  M. Haas
  \inst{1}
  \and 
  M. Ramolla
  \inst{1}
  \and
  C. Bruckmann         
  \inst{1}
  \and
  C. Westhues          
  \inst{1}
  \and
  R. Chini
  \inst{1,2}
  \and 
  K. Steenbrugge
  \inst{2,3}		
  \and
  R. Lemke
  \inst{1}
  \and
  M. Murphy
  \inst{4}
  \and
  W. Kollatschny
  \inst{5}		
}
\institute{
  Astronomisches Institut, Ruhr--Universit\"at Bochum,
  Universit\"atsstra{\ss}e 150, 44801 Bochum, Germany
  \and
  Instituto de Astronom\'{i}a, Universidad Cat\'{o}lica del
  Norte, Avenida Angamos 0610, Casilla
  1280 Antofagasta, Chile
  \and
  Department of Physics, 
  University of Oxford, 
  Keble Road,
  Oxford OX1 3RH, UK
  \and
  Departamento de F\'{i}sica, Universidad Cat\'{o}lica del
  Norte, Avenida Angamos 0610, Casilla
  1280 Antofagasta, Chile
  \and
  Institut f\"ur Astrophysik, Universit\"at G\"ottingen, 
  Friedrich-Hund-Platz 1, 37077 G\"ottingen, Germany
}

\authorrunning{F. Pozo Nu\~nez et al.}

\date{Received ; accepted}

\abstract{
  We consider photometric reverberation mapping, where the nuclear
  continuum variations are  
  monitored via a broad-band filter and the echo of emission line
  clouds of the broad-line region 
  (BLR) is measured with a suitable narrow-band (NB) filter. We
  investigate how an incomplete emission-line coverage by the NB filter influences the BLR size
  determination. This includes two basic cases: 1) a  
  symmetric cut of the blue and red part of the line wings, and 2) the
  filter positioned asymmetrically  
  to the line centre so that essentially a complete half of the
  emission line is contained in the NB filter.  
  Under the assumption that the BLR is dominated by circular Keplerian
  orbits, 
  we find that symmetric cutting of line wings may lead to overestimating
  the BLR size by less than 5\%. The case of
  asymmetric half-line coverage, similar as for our data of the Seyfert 1 galaxy
  3C120, yields a BLR size with a bias of less than 1\%. Our results
  suggest that any BLR size bias due to a narrow-band line cut in
  photometric reverberation mapping is small and in most cases
  negligible. 
  We used well-sampled photometric reverberation mapping light curves
  with sharp variation features in both the continuum and the H$\beta$
  light curves to determine the geometry
  type of the H$\beta$ BLR for 3C120. 
  Modelling of the light curve, under the assumption that the BLR is
  essentially virialised, argues against a spherical geometry and
  favours a
  nearly face-on disk-like geometry with an inclination $i = 10^\circ \pm 4^\circ$
  and an extension from 22 to 28 light days. 
  The low inclination may lead to a larger black hole mass 
  $M_{\rm BH}$ than
  that derived when using the average geometry scaling factor $f=5.5$. 
  We discuss deviations of Seyfert 1 galaxies from the 
  $M_{\rm BH} - \sigma_{*}$ relation.
} 

\maketitle
%

\section{Introduction}

The broad-line region (BLR) of active galactic nuclei (AGNs) has been
studied for over 40 years. In our current understanding of the
structure of the central engine of AGNs, the observed broad emission
lines are photo-ionised by the central ionising source located close to
the super-massive black hole (SMBH), which emits high-energy photons as a
consequence of matter falling into the SMBH. The broad emission lines
respond to the strong and variable optical/UV continuum at remarkably
short time-scales, and are thus at small distances (about 1 to 250
light days) from the accretion disk (AD) (e.g., \citealt{2000ApJ...533..631K};
\citealt{2009ApJ...697..160B}). For a virialised BLR with known geometry the mass
$M_{BH}$ of the SMBH can be determined.  

Reverberation mapping (RM, \citealt{1972ATsir.688....1L}; \citealt{1973ApL....13..165C}; 
\citealt{1982ApJ...255..419B}; \citealt{1986ApJ...305..175G}; \citealt{1993PASP..105..247P}) allows us, independent of spatial resolution, to infer the size
and morphology of the BLR.  
Spectroscopic monitoring measures the time delay
($\tau_{BLR}=R_{BLR}/c$) between changes in the optical-ultraviolet
continuum produced in the compact AD and the emission line from the BLR
gas clouds farther out. If the BLR clouds are essentially in virialised motion
around the black hole (e.g., \citealt{1999ApJ...521L..95P}; \citealt{2001ApJ...551...72K};
\citealt{2002ApJ...572..746O}; \citealt{2003A&A...407..461K}), combining the velocity
dispersion of the BLR gas ($\sigma_{V}$) with the BLR size ($R_{BLR}=c
\cdot \tau_{BLR}$) allows us to estimate the virial BH mass using
${M_{BH}} = f \cdot R_{BLR} \cdot \sigma_{V}^2/{G}$, where $G$ is the
gravitational constant, and the factor $f$ depends on the geometry and
kinematics of the BLR (\citealt{2004ApJ...613..682P} and references
therein).
The commonly used average factor $f$ of a small sample of AGN 
(\citealt{2004ApJ...615..645O}) may lead to a large scatter in ${M_{BH}}$ and in
relationships  
involving ${M_{BH}}$ (\citealt{2004ApJ...613..682P}). To improve
these relationships one needs  
to determine $f$ more precisely for each AGN.

Geometric models for spherical inflow, outflow, and disk BLR structures
have been constructed by interpreting the RM transfer function as a
two-dimensional map, the well-known "echo image" of the BLR (\citealt{1991ApJ...379..586W}; \citealt{2003SPIE.4854..262H}). Although there is no clear
consensus about the geometry of the BLR, past studies have shown
evidence in favour of a disk-like configuration (e.g., \citealt{1986ApJ...302...56W};
\citealt{1997ApJ...474...91M}; \citealt{1997ApJ...479..200B}; \citealt{2000ApJ...545...63E}; \citealt{1994ApJ...434..446K}; 
Kollatschny 2003a, 2003b, Kollatschny \& Zetzl 2011, 2013). 
Kollatschny \& Zetzl (2011, 2013) analysed the shape of the rms line
profiles of a homogenous AGN subsample of \citet{2004ApJ...613..682P} 
and found evidence for the superposition of rotational disk-like 
components and turbulence.

The velocity-resolved (two-dimensional) RM of Mrk 110 (Kollatschny
2003a) has best been 
explained with a dominating disk-like BLR component at inclination
$\sim$20$^{\circ}$ 
(i.e. close to face-on), consistent with the detected gravitational redshift
(Kollatschny 2003b). By using Bayesian probability theory and Monte Carlo algorithms,
\citet{2011ApJ...730..139P} 
presented a general method for geometric and dynamical
modelling of the BLR. Using this method, \citet{2011ApJ...733L..33B} and
\citet{2012ApJ...754...49P} modelled the geometry and estimated the BH mass for
the Seyfert 1 galaxies Arp 151 and Mrk 50. In both studies evidence
for a disk-like BLR geometry was found. 
The broad-line radio galaxy 3C120 has been observed with velocity-resolved RM 
at two different epochs (\citealt{2012ApJ...755...60G} and
\citealt{2014A&A...566A.106K}). Below we discuss details of this when we
compare their results with our data (Sect. \ref{section_data}).

In general terms, any modelling to infer the BLR geometry requires
high time-resolution 
observations and pronounced well-defined variability features in the light
curves. Spectrophotometric light curves are to date the main
source of observational data for RM. 
So far, velocity-resolved RM has been the only method used to study
the geometry of the BLR with high precision. Because the method is
resource expensive, it is desirable to explore whether the basic
features of the BLR geometry can be inferred directly from
photometric monitoring data.
 
Recently, photometric reverberation mapping (PRM) has been revisited
as an efficient method for determining the BLR size by
combining narrow- and broad-band filters (\citealt{2011A&A...535A..73H}; \citealt{2012A&A...545A..84P}, 2013) or 
by using only broad-band filters (\citealt{2012ApJ...747...62C}; \citealt{2012ApJ...756...73E}; \citealt{2012ApJ...750L..43C}; \citealt{2013ApJ...772....9C}; \citealt{2013ApJ...769..124C}). PRM may become an extremely powerful tool with unprecedented
efficiency -- in particular with respect to upcoming large surveys
like the LSST. Before the PRM is widely used, however, the accuracy
and possible biases have to be checked and established. Several such
studies are under way.

By using well-sampled light curves obtained for 3C120, we demonstrated
that the H$\beta$ BLR size R(H$\beta$) obtained by PRM is consistent
with that from spectroscopic RM (\citealt{2012A&A...545A..84P}, hereafter 
called Paper I). Similarly, with PRM we measured R(H$\alpha$) of the narrow-line
Seyfert 1 galaxy ESO399-IG20 (\citealt{2013A&A...552A...1P}, hereafter
called Paper II). It places ESO399-IG20 extremely well within the
$R-L$ relationship (refined by \citealt{2009ApJ...697..160B}), which supports the
correctness of the BLR size as derived from PRM. The strong H$\alpha$ line
contribution to the $r$ band allows us to derive R(H$\alpha$) even with
the pure broad-band PRM techniques proposed by \citet{2012ApJ...747...62C}. Furthermore, our PRM light curves of ESO399-IG20 show sharp
variation features in both the triggering continuum and the emission
line echo, enabling us -- by means of simple BLR models -- to find
evidence in favour of a nearly face-on thin disk-like BLR geometry.

In Papers I and II we monitored
the nuclear continuum variations via a broad-band (BB) filter, and the echo
of emission line clouds of the BLR was measured with a suitable narrow-band (NB) filter. However, sometimes the NB filter is too
narrow and thus cuts parts of the emission line (band width
$\sim$50\,\AA~ corresponding to a velocity coverage of
$\sim$3000\,km/s). The line wings correspond to BLR clouds with
the fastest line of sight velocities. Assuming Keplerian motion, these clouds
populate the innermost zones of the 
BLR. Thus, for photometric RM, one may expect that cutting the line
wings may result in an overestimate of the BLR size. Another
configuration is that the line is not centred on the NB
filter. Another difficulty, related to incomplete line coverage due to the NB is to
probe which conclusions about the BLR geometry can be drawn from such
PRM data with truncated velocity information.

This paper deals with three topics: 
in Sect.~\ref{section_modelling} we investigate how an incomplete emission line coverage
by the NB filter influences the BLR size 
determination. Specifically, we examine the bias
of the BLR size 1) for a symmetric cut
of the blue and red part of the line wings 
(Sect.~\ref{section_symmetric}), and 2) if the filter is
positioned asymmetrically to the line centre so that essentially a
complete half of the emission line is contained in the NB filter
(Sect.~\ref{section_asymmetric}). 
In Sect.~\ref{section_blr_geometry}, invoking reasonable assumptions from velocity resolved 
RM of 3C120, we apply simple BLR models to the well-sampled PRM data
of 3C120. In this case, the sharp variation features in the PRM data
enable us to constrain the BLR geometry of 3C120. 
Since a nearly face-on disk-like BLR may lead to a larger black
hole mass, we discuss the implications for the 
$M_{\rm BH} - \sigma_{*}$ relation in Sect.~\ref{section_mbh}.

\begin{figure} 
  \centering
  \includegraphics[width=\columnwidth]{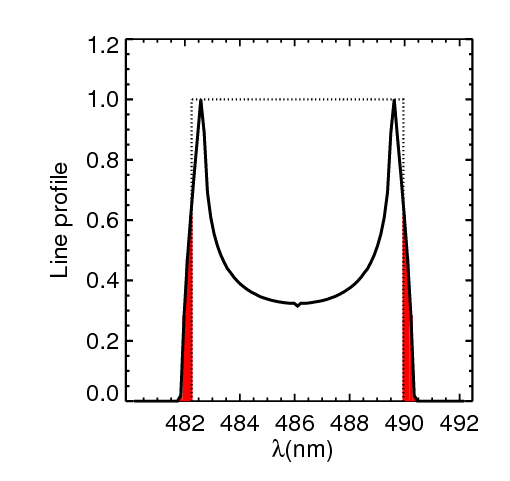}
  \includegraphics[width=\columnwidth]{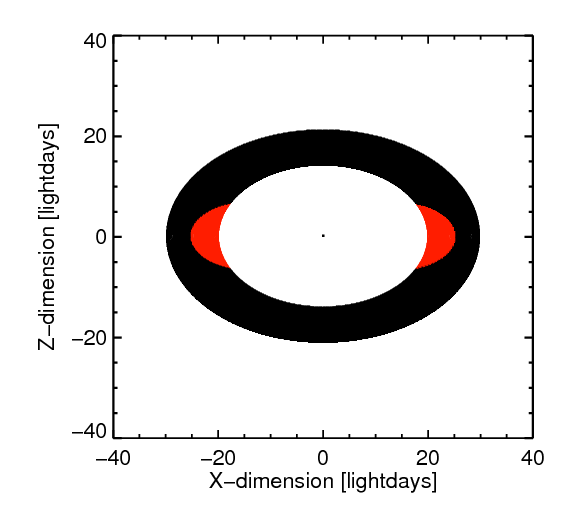}
  \caption{Symmetric case of a thin disk-like BLR geometry ($i =
    45^\circ$, $R_{min}$ = $20$ light days, $R_{max} = 30$ light days,
    $v_{max}$ = $3700$ km/s). Top: a rectangular filter (dotted line)
    is symmetrically centred on the spectral line (thick solid line);
    the parts of the line that are cut off by the filter are shaded
    in red. Bottom: BLR areas inside (black)  
    and outside (red) of the filter. If a disk-like BLR-model with
    Keplerian orbits is 
    assumed, the red shaded areas with the highest apparent velocity
    are cut off by the filter.
  } 
  \label{fig_symm_filter_and_ring}
\end{figure} 

\section{Data and model considerations}
\label{section_data}
 
The source 3C120 has been monitored spectroscopically -- assisted by BB
photometry -- in 2008/2009 by Kollatschny
et al. (2014, K+2014) and in 2010/2011 similarly by Grier et
al. (2012, 2013, G+2013). 
The data by G+2013 are very well sampled, but catch only
rather smooth variations in the light curves. 
The data by K+2014 show strong variations, but are not
optimally sampled (median 6-8 days).
Both G+2013 and K+2014 performed a velocity-resolved RM analysis, albeit
with different techniques, and obtained mostly consistent but slightly
different results. 
Both analyses found evidence for a disk-like BLR component, but 
G+2013 interpreted the red
wing in the H$\beta$ and H$\gamma$ lines as signature for inflowing gas,
while K+2014 interpreted the red Balmer wings as optical-depth
effects of a disk-wind (as proposed by \citealt{1997ApJ...474...91M}). Moreover, from the He lines,
K+2014 found evidence for outflows, that appear to be more consistent with
the expectations for this powerful broad-line radio galaxy having a
superluminal jet pointing closely towards us (see Sect.~\ref{section_results}).  

The photometric RM data of 3C120 that are examined here were observed in
2009/2010 (Paper I), between the two spectroscopic RM campaigns,
when 3C120 was in a lower luminosity state. In brief, we observed
light curves in the $B$ band, the [OIII] 5007\AA~ NB
to catch the $z = 0.033$ redshifted H$\beta$ line, 
and the $V$ band to subtract host
and continuum contribution from the H$\beta$ light curve.
Our well-sampled PRM light curves (median two days) show 
strong abrupt variations in both 
continuum and H$\beta$, but the velocity information is missing and
the velocity range is truncated. 

Before we describe the model in detail, we stress two general consideration.

Firstly, if the BLR has a spherical geometry, the front and back sides of
the sphere will broaden the transfer function and thus will 
smear out the echo, which then looses its sharp features.
The same holds for a disk-like BLR with high inclination $i > 40^{\circ}$
(i.e. intermediate to edge-on view).
Because this smearing-out contradicts what is observed in our echo
light curves of 3C120,
any spherical BLR or highly inclined BLR disk appears to be unlikely.
This conclusion is independent of the kinematical model. 
In more general terms, the direct visual inspection of the light
curves suggests that the sharp H$\beta$ echoes, which perfectly follow the sharp continuum
variations, exclude any BLR geometry, that is extended along
the line of sight. 

Secondly, if the kinematic model of the BLR
(Kepler movement of gas clouds or outflow) is known, for instance from
velocity--delay maps, and if the BLR is
quasi-static during the observing campaign, the echo light curve
solely depends on the location of the echoing gas clouds. 
This is why PRM data are able to provide
constraints on the geometry. 

We modelled the photometric RM data of 3C120 in essentially the same
way as we did for ESO399-IG20 in Paper II. The models are below characterised briefly.

Following Welsh \& Horne (1991), we considered simple models.  
We assumed circular Keplerian orbits,
thin disks, and/or spherical BLR geometry, 
with inner and outer radii ($R_{min}$, $R_{max}$).
Consistent with the optimally emitting cloud model (\citealt{1995ApJ...455L.119B}), we found and assumed 
modest ratios $R_{max}$ / $R_{min} < 2$, and therefore used throughout a constant
gas density independent of the cloud distance from the very centre,
and a constant radiation efficiency $\epsilon$. 

We varied the thin-disk inclination $i$, and finally also used modestly
thick inclined disks with covering 
half angles $\alpha$. Our simple models do not account for
in-/outflow, turbulence, projection effects, 
or for optical-depth effects in the disk-wind model of \citet{1997ApJ...474...91M}.

\section{Effect of cutting the line wings on the measured mean lags}
\label{section_modelling}

Photometric reverberation mapping uses a NB filter to
observe the emission line and a BB filter  
to observe the continuum. Here we examine the possible bias in lag 
determination caused by NB filter line-wing cutting. Line wings
correspond to BLR clouds with the fastest 
line-of-sight velocities. The objective is to quantify the bias with
simple models and to check photometric lag estimations.  
We restricted our investigation to thin disk-like BLR geometries, considering them as worst cases,
because disks have a higher proportion of line wings (fast BLR clouds)
than spherical BLRs, hence disks are more affected by line-cutting.

\begin{figure} 
  \centering
  \includegraphics[width=\columnwidth]{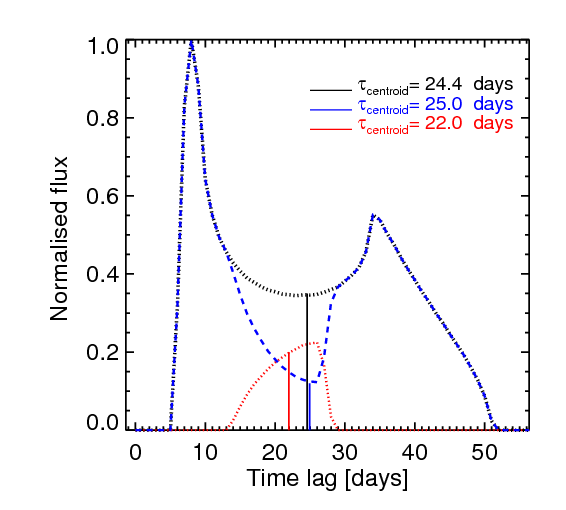}
  \caption{Response of the BLR
    to a single light-pulse. The BLR is modelled as a thin circular
    Keplerian-disk. The dotted
    curves show the time-delay function of the entire BLR (black), the
    BLR part outside of the NB (red dotted line), and the BLR part
    inside of the NB (blue dashed line). The response of the
    entire emission line 
    (black) can be separated into a part contained by the NB  
    and a part outside the NB. The solid vertical lines
    mark the centroids of the filter answers (i.e. the mean time
    lag). The difference between 
    the blue and the black centroid is the response (echo) bias. The
    value of the measured echo (blue) is higher than that of the real one
    (black). Thus, 
    cutting line wings will (marginally) increase the mean time lag.
  } 
  \label{fig_symm_lags}
\end{figure}

\subsection{Symmetric case for a rectangular band-pass}
\label{section_symmetric}

We start with an inclined disk-like BLR and assumed a
rectangular bandpass for the filter centred on the emission line, as
illustrated in Figure~\ref{fig_symm_filter_and_ring}. At first glance, one expects that the average echo of
these innermost clouds (with a velocity outside the band pass)
arrives earlier than the mean echo of the complete disk. 

The echo of
an isotropic light-pulse from the 
(point-like) central accretion disk was calculated as shown in
Fig.~\ref{fig_symm_lags}. The first 
echo peak lags the continuum pulse by about eight days. It originates from
the front side of the ring, which is tilted towards the observer. The latest echo from the back
side arrives about 40-50 days 
later than the continuum pulse. To compare this with the observations,
we measured the mean lag by the centroid of the time-delay function, which corresponds to the centroid of the cross-correlation of two light curves. The 
mean true lag of the entire BLR is 24.4 days. The BLR part missed by
the NB filter is that 
with the highest line-of-sight velocity, as shown in 
Fig.~\ref{fig_symm_lags}. The echo of this region has a mean lag of 22 
days. But for the BLR covered by the NB, the
measured echo has a mean lag of 
25 days. This is only 2-3\% more than the true mean BLR size. The
reason for this remarkably small 
deviation between measured and true BLR size is that the loss occurs close to the centroid of the response function, 
while the outer parts remain unaffected (Fig.~\ref{fig_symm_lags}).

\begin{figure} 
  \centering
  \includegraphics[width=\columnwidth]{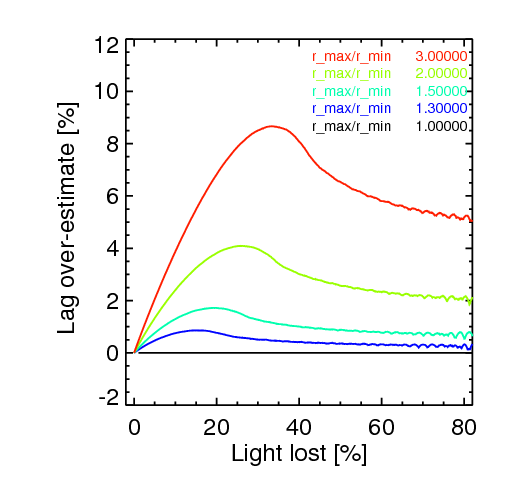}
  \caption{Dependence of the time-lag over-estimate 
    for a thin-disk BLR at inclination 45$^\circ$. 
    For each value of $r_{max}$/$r_{min}$, the curve 
    shows the lag over-estimate as a function of line cut, 
    here expressed as light lost.
  }
  \label{fig_symm_overestimate}
\end{figure}

We computed the effect of line cutting for a range of 
parameters: inclination $i$, $r_{min}$ and $r_{max}$, 
and $M_{\rm BH}$, i.e. $v_{max}$ at $r_{min}$. $v_{max}$ is the highest velocity seen at the given inclination $i$; in
reality $v_{max}^{real}$ = $v_{max}$ / sin ($i$). The most important parameter -- apart from the geometry 
type -- is the ratio of $r_{min}$ and $r_{max}$
of a BLR, virialised with gas clouds on circular Keplerian orbits, 
and the percentage of light that is cut by the NB filter. 
The resulting over-estimations of the time lag are shown in 
Fig.~\ref{fig_symm_overestimate}. 
For typical cases $r_{max}/r_{min} < 2$; 
if the NB cuts off less than 10\% (20\%) of the line
flux, the lag over-estimate is lower than 2\% ($\sim$4\%). 

\begin{figure} 
  \centering
  \includegraphics[width=\columnwidth]{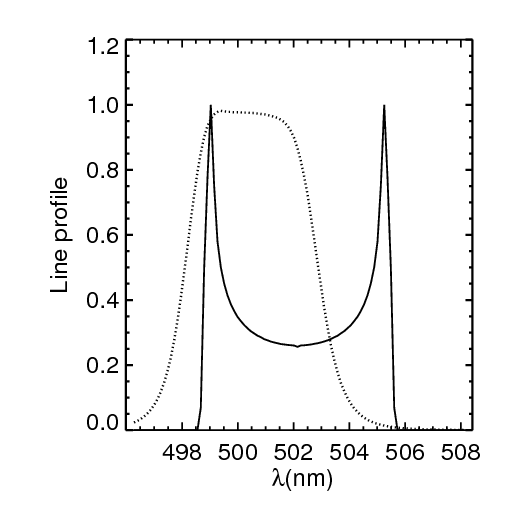}
  \includegraphics[width=\columnwidth]{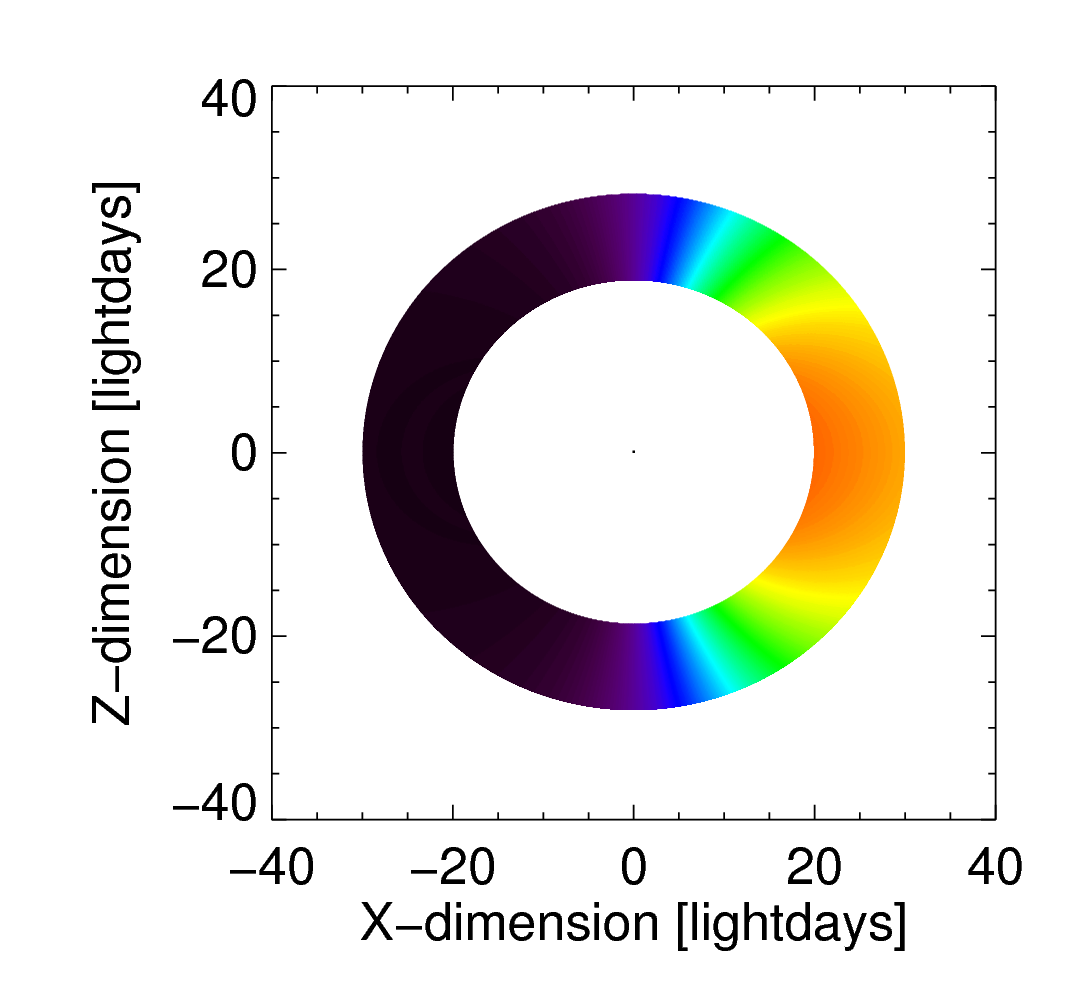}
  \caption{
    Asymmetric case for a disk-like BLR such as 3C120 ($i =
    10^\circ$, $R_{min}$ = $22$ light  
    days, $R_{max}$ =$28$ light days, 
    $v_{max}$ = $2850$ km/s). 
    Top: 
    NB filter ([OIII]$\lambda$ 5007\AA, dotted line) 
    and emission-line profile of the BLR (H$\beta$, thick solid line).
    The filter essentially covers the complete blue half 
    of the H$\beta$ line, as is the case for 3C120. 
    Bottom: 
    BLR areas inside (black) and outside (coloured) 
    of the NB filter.
    The colour-coding (blue, green, red) 
    indicates a line  transmission of about 50\%, 20\%, and 0\%. 
  }
  \label{fig_asym_filter_and_ring}
\end{figure}

\subsection{Asymmetric case for 3C120}
\label{section_asymmetric}

The symmetric case of a line that is centred on the NB is
the exception not the 
rule. The question is whether an asymmetrically positioned filter would
exacerbate the situation. 
3C120 lies at redshift $z = 0.033$ so that the wavelength of the
H$\beta$ line is shifted from $486.1$ nm to $502.1$ nm. 
Thus, H$\beta$ falls roughly within the [OIII] filter ($500.7$ nm, FWHM
$5$ nm), but not symmetrically 
or completely, as shown in Figure~7 of Paper I. The H$\beta$ line has
a quite 
symmetric profile, and nearly the complete blue half of the line falls
within the [OIII] NB. 

To illustrate the effects of this line/filter mismatch, we used the
parameters for the BLR 
of 3C120 that were found to be the best solution in the simulation below
(Sect.~\ref{section_blr_geometry}), adopting circular Keplerian 
orbits. Figure~\ref{fig_asym_filter_and_ring} shows the part of the
BLR that is entirely covered by the NB 
filter and the parts that are missed or only partly contained in the NB filter. 

\begin{figure} 
  \centering
  \includegraphics[width=\columnwidth]{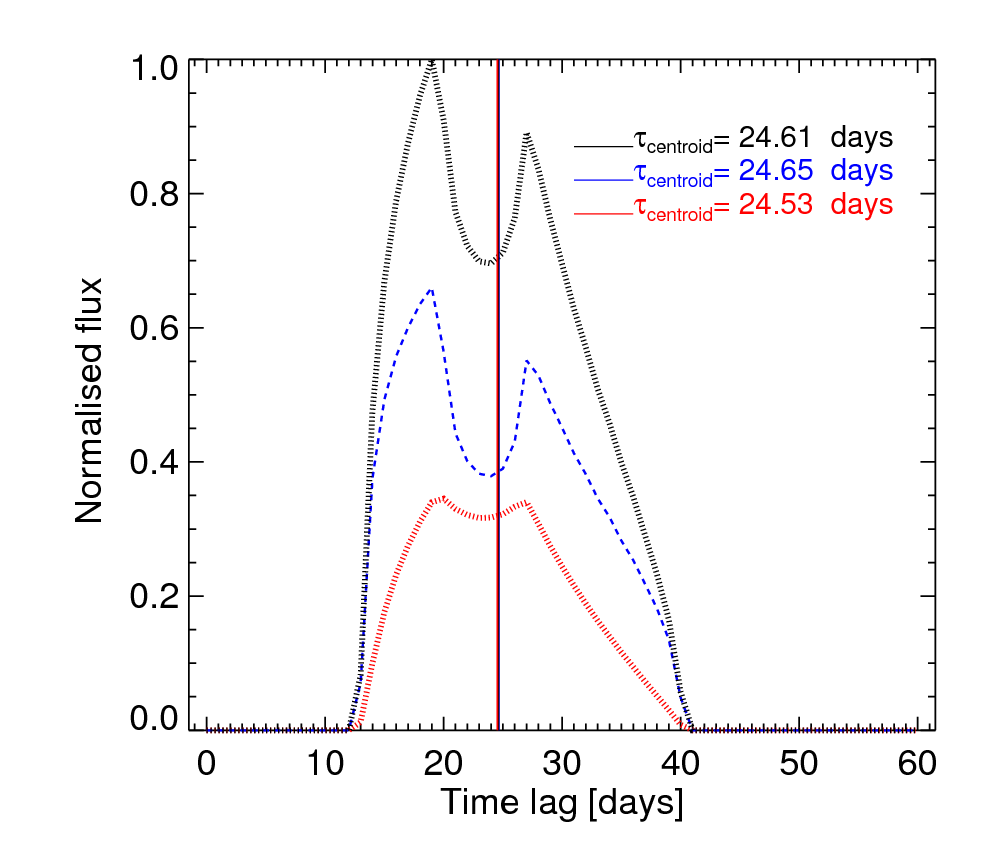}
  \caption{Asymmetric case for a disk-like BLR such as 3C120 ($i =
    10^\circ$, $R_{min}$ = $22$ light days, $R_{max} =28$ light days,
    $v_{max}$ = $2850$ km/s). The dotted lines show the time-delay 
    function of the entire BLR (black), the BLR part outside of the
    filter (red), and the BLR part inside the filter (blue). The
    centroids of the respective curves are plotted by the vertical
    solid lines, which virtually coincide.} 
  \label{fig_asym_lags}
\end{figure}

Figure~\ref{fig_asym_lags} shows the echo of an isotropic light-pulse
from the 
(point-like) central 
accretion disk.  
The curves stand for the entire BLR, the missing part of the
BLR (less than 20\% transmission) 
and the part contained in the NB filter (more than 50\%
transmission). The centroids of 
the light echoes agree nicely, demonstrating that any bias in the lag
determination is negligible -- under the premise that the chosen BLR model,
i.e. a thin circular Keplerian disk, is correct.  

\section{BLR geometry of 3C120 from light curve modelling}
\label{section_blr_geometry}

The aim is to distinguish between a spherical and a disk-like
BLR geometry by means of light curve modelling. 
Figure~\ref{fig_detrending} (top) shows the $B$ band and H$\beta$ light
curves (derived using 7$\farcs$5 apertures).

The light 
curves are well sampled and rich in features. 
The $B$ band light curve was corrected for host-galaxy
contributions (see Paper I), and hence essentially represents 
the pure AGN light curve, which was used in the modelling as triggering input
for the simulations. The H$\beta$ light curve was 
constructed from the [OIII] NB light curve by essentially correcting
via the $V$ band light curve for continuum contributions
(Paper I). This $V$ band correction may be not accurate enough, and thus 
the resulting H$\beta$ light curve may still
contain non-variable contributions, for instance 
from the narrow-line region. To use this H$\beta$ light curve
as a tracer for the BLR echo, these 
contributions need to be removed (Sect.~\ref{section_detrending}). 
The reason is that otherwise the amplitude of the modelled echo
becomes always too large compared with the amplitude of the observed
echo, which prevents a reliable model fit. 
Another concern is the gap in observational data points between days 75
and 95, which we consider in Sect.~\ref{section_spear}.

\subsection{Correcting the H$\beta$ light curve for non-variable contributions}
\label{section_detrending}

\begin{figure} 
  \centering
  \includegraphics[width=\columnwidth]{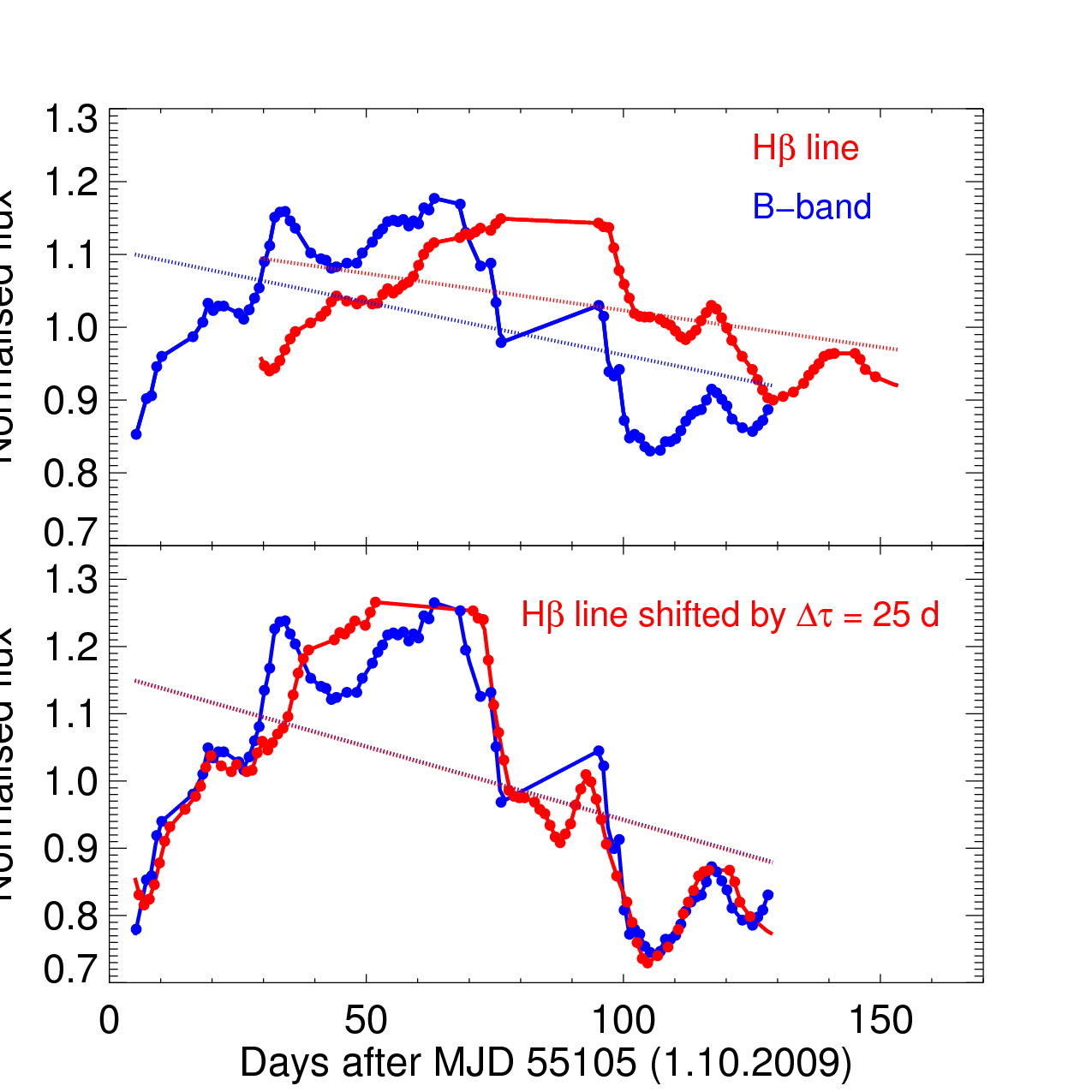}
  \caption{Light curves of 3C120 in $B$ (blue) and 
    H$\beta$ (red), normalised to their mean. 
    Top: The thin dotted lines represent the long-term trend
    quantified 
    by a linear fit to the light curves. 
    The long-term trends of $B$ and H$\beta$ appear to
    differ because a non-variable flux contribution 
    in the H$\beta$ light curve. 
    Bottom: after determining and
    subtracting the non-variable flux in the 
    H$\beta$ light curve ($C\sim 25 \%$) and re-normalisation, the long-term trends
    agree. 
    For visual purposes we shifted the H$\beta$ light curve by an
    estimated lag time of 25 days.  
  }
  \label{fig_detrending}
\end{figure}

To determine the potential non-variable contribution to the H$\beta$
light curve, we performed a long-term trend analysis. The BLR
size of $\sim$25 light 
days is smaller than the duration of the monitoring campaign. 
Thus, on the long time-scale ($\sim$150 days) both $B$ band and
H$\beta$ light curves should show the same long-term trend, but this is not the case in the original data (Fig.~\ref{fig_detrending},
top), which indicates an additional contribution to the
H$\beta$ light curve. After subtracting a suitably scaled
non-variable contribution   
(C$\sim$25\%) from the H$\beta$ light curve and subsequently
re-normalising the long-term trends agree 
(Fig.~\ref{fig_detrending}, bottom).

\begin{figure} 
  \centering
  \includegraphics[width=\columnwidth]{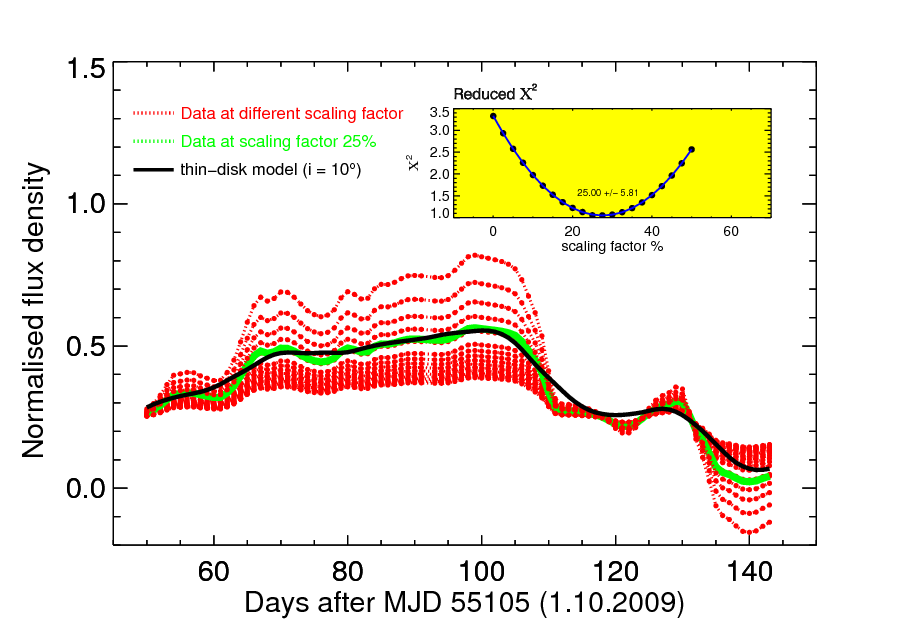}
  \caption{
    H$\beta$ light curves of 3C120 with different non-variable
    contributions C between 0\% and 50\% subtracted and re-normalised (red).
    The best model fit is shown in black and the best corresponding
    data in green. The inset shows the reduced chi-square as a function
    of $C$. The best chi-square is for $C\sim 25\%$, perfectly consistent with
    the $C$-value from the long-term trend analysis in Fig.~\ref{fig_detrending}.
  }
  \label{fig_remove_continuum}
\end{figure}

Because the transfer function is not a $\delta$-function, but
spread over $\sim$25 days, the time span of 150 days
might be too short to apply the long-term trend correction above. 
Therefore, we derived the correction factor C of the non-variable
contribution to the H$\beta$ light curve in an alternative way. For a
grid of subtracted constant contributions (C between 0\% and 50\%) we
applied the model fitting and compared the reduced chi-square
values. The best fits were obtained for
C$\sim$25\% (Fig.~\ref{fig_remove_continuum}).

The perfect agreement of C$\sim$25\% derived by both the long-term
trend and the fitting process gives us confidence that any
non-variable continuum in the H$\beta$ light curve is well
corrected for.  
Thus, we are confident to have obtained the proper amplitude scaling
between the triggering 
$B$ band variations and those of the H$\beta$ echo.

\subsection{Applying SPEAR to bridge the observational gap}
\label{section_spear}
 
Despite the dense sampling of two days (median over six months), the
observed PRM light curves have a gap of 20 days between December
2009 and January 2010 (days 75 to 95 in Fig.~\ref{fig_detrending},
top). These gaps 
constitute about 20/150 $\sim$14\%, i.e. a small fraction, of the
monitoring. So far, we have linearly interpolated the gaps and
obtained good fit results.

\begin{figure} 
  \centering
  \includegraphics[width=\columnwidth]{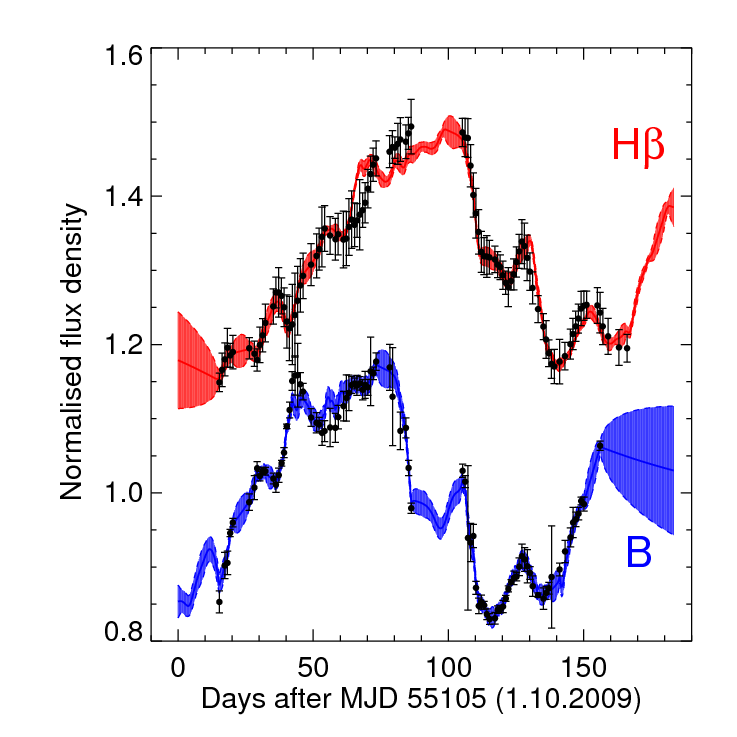}
  \caption{
    Interpolated continuum and H$\beta$ light curves. 
    The solid red and blue lines show the H$\beta$
    and continuum models estimated by SPEAR. 
    The red and blue areas (enclosed by the dashed line)
    represent the expected variance about the mean light 
    curve model obtained with SPEAR. 
    The H$\beta$ light curve (black dots) is vertically shifted 
    with respect to the continuum light curve (black dots)  
    for clarity.
  }  
  \label{fig_spear}
\end{figure}

On the other hand, linear interpolation may have missed important
variation features during the gaps. It has be shown that the AGN
continuum variability can be reproduced by a damped random walk
(DRW) process (\citealt{1992ApJ...391..518V}; \citealt{2009ApJ...698..895K}; \citealt{2012ApJ...753..106M}). To check for consistency with stochastic models, we
interpolated the continuum applying the code stochastic process
estimation for AGN reverberation (SPEAR) (\citealt{2011ApJ...735...80Z}) --
similar to what we did in Paper II.  
Figure~\ref{fig_spear} shows the light curve models obtained with
SPEAR. They do not differ much from the interpolated light curves
(Fig.~\ref{fig_detrending}, top). 
In the gaps the difference between the SPEAR and the interpolated
light curves is lower than 5\% and $\sim$7 times lower than the
overall variability amplitues (35\%).
Because of the small difference
between the data and the SPEAR model light curves, and to avoid to
fitting a BLR model to a light curve that itself is only a modelled
curve, we decided to use the interpolated data light curves for
the subsequent modelling of the BLR.

\subsection{Results and discussion}
\label{section_results}

\begin{figure} 
  \centering
  \includegraphics[width=\columnwidth]{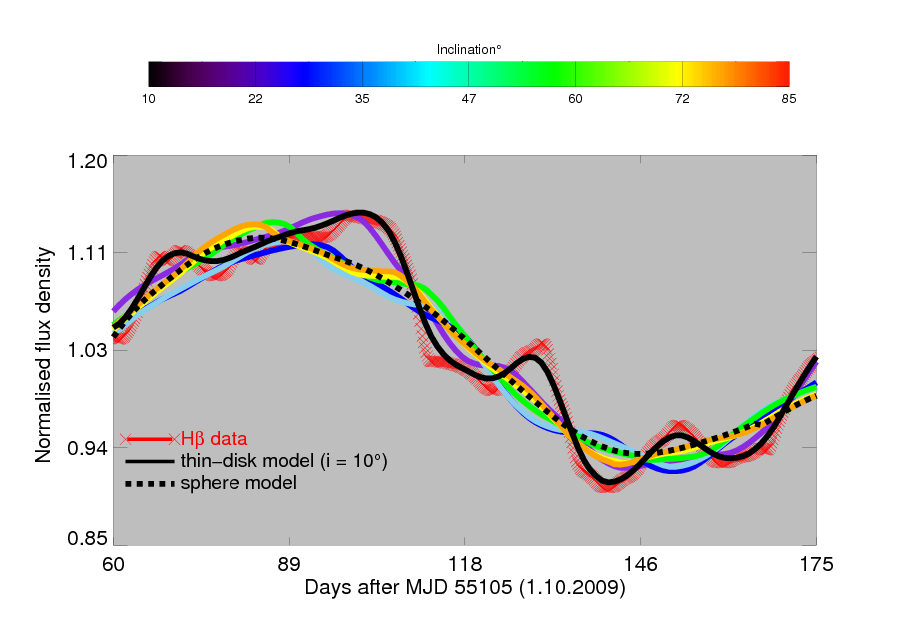}
  \caption{ H$\beta$ and best-fit model light curves of 3C120.
      The interpolated H$\beta$ light curve
      is shown with red
      crosses. 
      Note the sharp drop of the light curve at days $\sim$110 and
      $\sim$130 as well as the sharp feature at days $\sim$150 and the
      strong rise at day $\sim$170.
      Both the spherical BLR model (black dotted line) and the
      inclined thin disks ($i=45^{\circ}$ blue solid line, $i=60^{\circ}$ green
      solid line) fail in fitting
      these sharp variation features. 
      The features are best fitted by 
      a thin-disk BLR model with inclination $i=10^{\circ}$ (black
      solid line). 
      Already at $i=20^{\circ}$ (violet solid line) the thin disk
      fit yields a poorer chi-square. 
      The light curves are shown only after the initial
      settling-down phase ($\sim$60 days), to account for the unknown
      continuum light curve before day zero.
    } 
  \label{fig_best_lc_fit}
\end{figure}

We used the host-subtracted $B$ band light curve as the one that trigger AGN continuum
light curve. We 
created echo model light curves, by convolving 
the AGN continuum light curve with the time-delay function for a spherical BLR
and for thin Kepler disks at a range of inclinations.
We compared the echo model light curves with the H$\beta$ light
curve. Figure~\ref{fig_best_lc_fit} shows the simulation results. Particularly important are the two pronounced sharp features in the
H$\beta$ light curve, for instance the sudden declines at days
$\sim$110 and $\sim$130. Even the best spherical fit does not reproduce these sharp features well, which makes a spherical BLR
for 3C120 highly unlikely. The same holds for the strongly inclined thin disks
$i > 40^{\circ}$. The best fit is a thin-disk BLR model with an
inclination $i= 10^{\circ}$, which well reproduces
the features of the H$\beta$ light curve; this provides strong
evidence in favour of a disk-like 
BLR in 3C120. From the chi-square analysis we obtain an inclination of
$i=10^{\circ}\pm4^{\circ}$ for a thin-disk BLR model with an extension
from 22 to 28 light days.  

\begin{figure} 
  \centering
  \includegraphics[angle=0, width=\columnwidth]{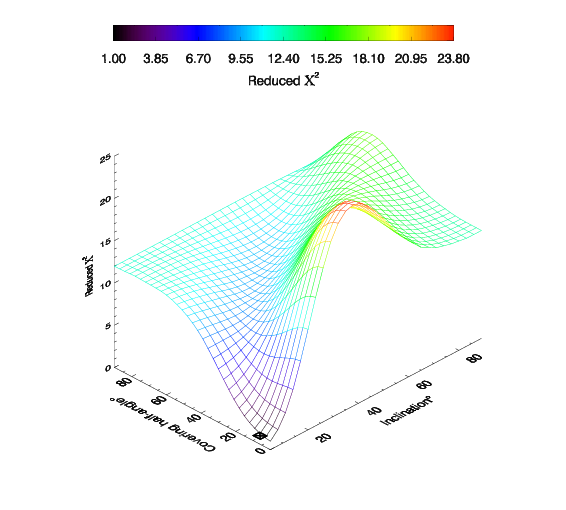}
  \caption{
    Distribution of reduced $\chi^2$ values for inclination $i$ and covering half-angle $\alpha$ of disk-like BLR models for 3C120. 
    The inclination starts at $i = 5^{\circ}$ 
    (and not at $i = 0^{\circ}$), and the grid is in steps of
    3$^{\circ}$ for both $i$ and $\alpha$.
    The black crossed square marks the place of the minimum $\chi^2$
    found at  $i = 5^{\circ}$ and  $\alpha = 10^{\circ}$.
    The inner and outer radius of the
    disks is 22 days and 28 days, respectively.
  }
  \label{fig_chi_map}
\end{figure}

As a refinement of the thin disks, we also modelled thick-disk
geometries with a range of covering angles. Our thick disks are
simply the superposition of several disks that cover a range of
inclinations. 
For instance, a thick disk with mean $i= 10^{\circ}$ and a covering
half-angle $\alpha = 10^{\circ}$ is composed of thin disks with
$0^{\circ} <i < 20^{\circ}$. The best chi-square of the thick-disk
model yields the same extent from 22 to 28 light days as the thin
disk model, but a slightly different inclination of
$i=5^{\circ}\pm6^{\circ}$, and a covering half-angle $\alpha =
10^{\circ}\pm15^{\circ}$.  
The chi-square performance for $i$ and $\alpha$ is shown in
Fig.~\ref{fig_chi_map}.
To estimate the $i$ and $\alpha$ uncertainties  
$\sigma_{i} = 6^{\circ}$ and $\sigma_{\alpha} = 15^{\circ}$ 
we used the chi-square contours at
$\chi^2 = 5$, i.e. at five times the minimum chi-square
values. 

The value $\sigma_{\alpha} = 15^{\circ}$ appears to be relatively high
and, for thick disks, rates an orbit at inclination $i=+20^{\circ}$ as
good, while it is rejected for a thin-disk fit (Fig.~\ref{fig_best_lc_fit}). 
We suggest that the relatively high $\sigma_{\alpha} = 15^{\circ}$
arises because the fits of our thick disks become degenerate for nearly
face-on orbits. For instance, a thick disk at mean $i=+10^{\circ}$ and 
$\alpha = 20^{\circ}$ contains -- among others -- two orbits one at  
$i=+30^{\circ}$ and one at $i=-10^{\circ}$. 
Because $\chi^2$ does not distinguish between an orbit at
$i=+10^{\circ}$ and one at $i=-10^{\circ}$, the excellent $i=-10^{\circ}$
orbit will counterbalance the poor contribution of the $i=+30^{\circ}$
orbit.
Nevertheless, the results favour a nearly face-on and rather
thin disk as the best-fit solution for 3C120.

Proper-motion studies of the superluminal radio jet of 3C120
revealed that the highest apparent velocity is around $6c$ (\citealt{2001ApJ...561L.161G}, \citealt{2001ApJ...556..756W}), which allows a largest jet-viewing 
angle of 20$^{\circ}$ relative to the line of sight (\citealt{2002Natur.417..625M}). More recently, and through modelling Very Long Baseline
Array (VLBA) observations, \citet{2012ApJ...752...92A} estimated a jet-angle 
inclination of $\sim$16$^{\circ}$. If the jet axis is aligned with the
normal vector of both the accretion disk and the BLR disk (\citealt{1975ApJ...195L..65B}), the
inclination values obtained from radio data are nicely consistent 
with our BLR inclination.

By modelling detailed spectroscopic reverberation mapping and
velocity-delay maps of 3C120, albeit with softer light curve
features than seen in our data, \citet{2013ApJ...764...47G} and \citet{2014A&A...566A.106K} also found evidence
for a disk-like BLR, but did not specify the parameters. Remarkably,
our results were obtained without   
explicit velocity information, but under the assumption that the BLR
is essentially virialised. The light curves do not depend on
the velocity of the BLR clouds, but on the  
location of the clouds, hence the BLR shape, which is quasi-static on the 25-day time scale considered here. The
velocity information is then only needed to distinguish between a 
Keplerian rotation and other kinds of cloud movement such as in- or
outflow, or turbulence.  
This explains why it is possible -- under favourable conditions as in
our case of 3C120 -- to infer the BLR shape even from photometric
reverberation data.

\section{Central black hole mass and the geometry scaling factor for 3C120}
\label{section_mbh}

If 3C120 has a nearly face-on disk-like BLR geometry with
$i\sim10^\circ$, the geometry-scaling factor $f$ may be much
higher than the commonly used average value  $f = 5.5$ (\citealt{2004ApJ...615..645O}), which in consequence may result in a much larger black hole
mass $M_{BH}$. While it is not possible to determine $M_{BH}$ with
high precision from PRM
data and single-epoch spectroscopy alone, we here discuss the
implications of a 
face-on disk-like BLR geometry, which are of widespread relevance.

Spectroscopic reverberation mapping is the only way to provide reasonably accurate
$M_{BH}$ estimates for AGN via the virial theorem:
\begin{eqnarray}
  {M_{BH}} = f \frac{R \cdot \sigma_{V}^2}{G},
\end{eqnarray}
where $\sigma_{V}$ is the emission-line velocity dispersion (assuming
Keplerian orbits of the BLR clouds), $R=c \cdot \tau$ is the BLR size, and the factor $f$ depends on the -- so far unknown -- geometry and
kinematics of the BLR (\citealt{2004ApJ...613..682P} and references
therein). RM also enables rough $M_{BH}$ estimates from  single-epoch
spectra through the
use of the BLR size-luminosity $R_{\rm{BLR}} \propto
L_{\rm{AGN}}^{0.5}$ relationship (\citealt{1991vagn.conf..339K}; \citealt{1996ApJ...471L..75K}; \citealt{1999ApJ...526..579W}; \citealt{2008ApJ...673..703M}; \citealt{2006ApJ...641..689V}; \citealt{2011nlsg.confE..38V}). By assuming that AGNs and
quiescent galaxies follow the same $M_{BH}-\sigma_{*}$ relationship,
\citet{2004ApJ...615..645O} have shown that the scaling factor ($f$) has an
average value of $5.5$, if the velocity width is determined by
considering the second moment of the line profile ($\sigma_{V}$)
rather than the FWHM ($V_{FWHM}$). Subsequent studies (based on the
method established by \citealt{2004ApJ...615..645O}) have shown that the intrinsic
scatter of the scaling factor $f$ is about 0.4 dex (\citealt{2006A&A...456...75C}; \citealt{2010ApJ...716..269W}; \citealt{2010ApJ...709..937G}; \citealt{2011MNRAS.412.2211G}),
indicating that the scaling factor is a crucial source of
uncertainties in determining black hole masses through the RM method or
single-epoch spectra (\citealt{2010ApJ...716..269W}).  

A different approach for estimating $M_{BH}$ is by means of the
observed gravitational redshift of high-ionization emission lines
(\citealt{1977MNRAS.181P..89N}; \citealt{1982ApJ...263...79G}; \citealt{1985ApJ...292..164P}; \citealt{1990ApJ...350..512Z};
\citealt{2003A&A...412L..61K}). Certainly, this technique requires high spectral
resolution monitoring to detect the small redshift, for
instance from the rms spectra, that is from the variable portion of the
broad emission-lines. In the pioneering study of Mrk110, \citet{2003A&A...412L..61K} estimated a gravitational mass of $14 \pm 3 \cdot 10^{7}\,
M_{\odot}$, which is higher than the  $2.5 \pm 6 \cdot 10^{7}\,
M_{\odot}$ (Peterson et al. 2004) obtained by spectroscopic RM
assuming a scaling factor $f=5.5$. Kollatschny found evidence for a
disk-like BLR with inclination $i \sim 21^{\circ}$.  

\begin{figure} 
  \centering
  \includegraphics[angle=90, width=\columnwidth]{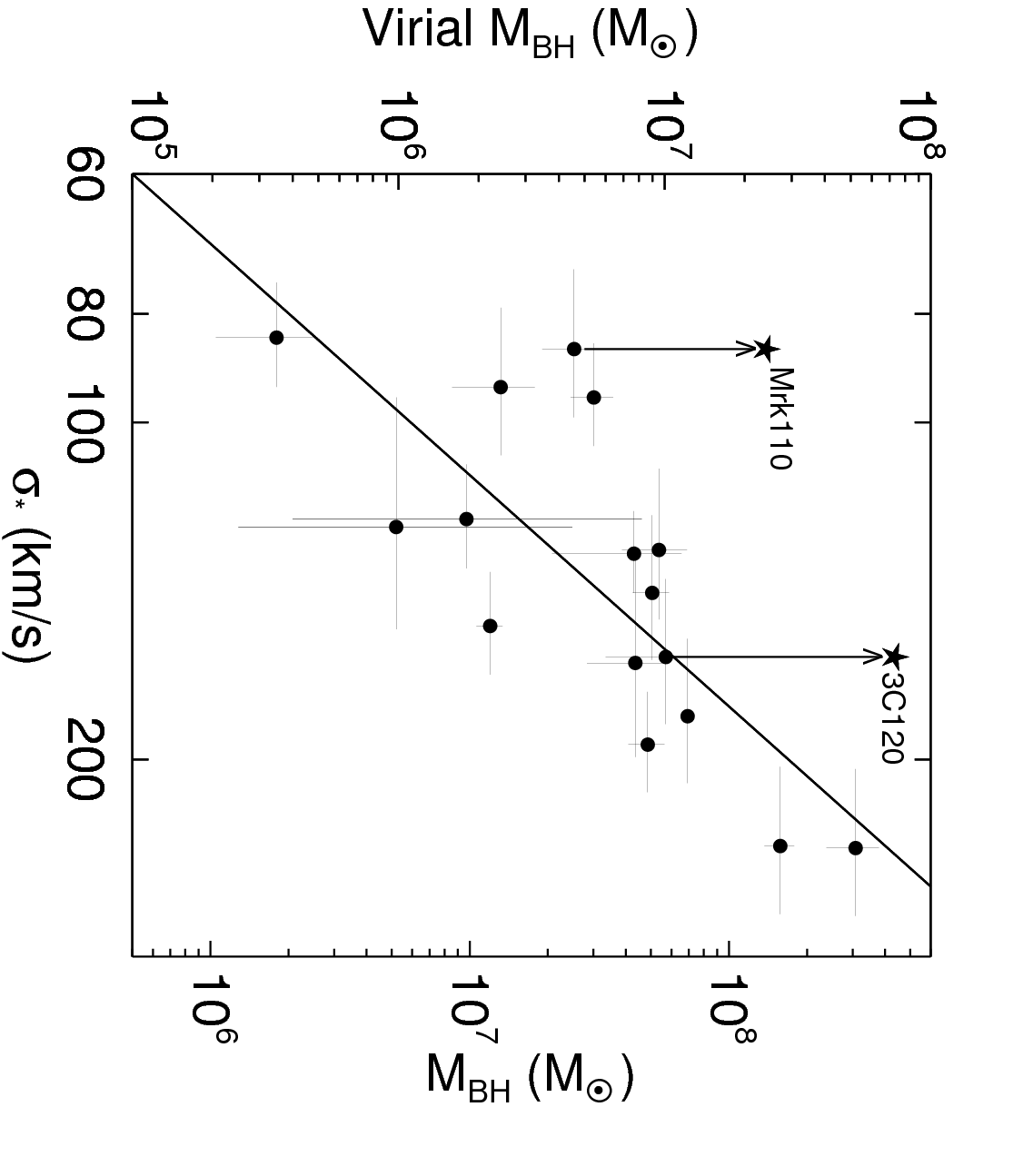}
  \caption{
    $M_{BH}$--$\sigma_{*}$ relationship from Onken et
    al. (2004) using data of Peterson et al. (2004)  (black dots). 
    The black arrows show the positional
    shift of 3C120 and Mrk110 with respect to the previous ones from
    Peterson et al. (2004). The solid line is the best-fit slope (case
    F02) from Onken et al. (2004).
}
  \label{fig_mbh}
\end{figure}

If 3C120 has a nearly face-on disk-like BLR geometry with
$i\sim10^\circ$, this results in a value $f=\frac{2 \cdot \ln
  2}{\sin^2{i}} = 46$ (ranging between 20 and 180 if the inclination
errors are taken into account). This value is about eight times higher
than the statistical value $f=5.5$  obtained by \citet{2004ApJ...615..645O}. 
For 3C120, \citet{2004ApJ...613..682P} determined 
a virial mass $M_{\rm virial} = M_{\rm BH}/f = 10.1 \pm 4 M_{\odot}$,
while \citet{2012ApJ...755...60G} found
$M_{\rm virial} = 12.2 \pm 1.2 \cdot 10^{6} M_{\odot}$. 
In Paper I
we estimated a virial mass $M_{\rm virial} = 10 \pm 5
\cdot 10^{6} M_{\odot}$. 
Using as a conservative approach the lowest $M_{\rm virial} = 10 \cdot 10^{6} M_{\odot}$ and the new derived scaling factor $f = 46$, 
the resulting black hole mass of 3C120 is
$M_{\rm BH} = 460 \cdot 10^{6}\, M_{\odot}$. 
Fig.~\ref{fig_mbh} shows the position of 3C120 in the
$M_{BH}-\sigma_{*}$ diagram. The data are taken from Onken
et al. (2004, their Table 3).
In all cases, $M_{\rm BH}$ has been calculated using 
the velocity dispersion (Eq. 4 of Peterson et al. 2004), 
and with the assumption that the line width is dominated by Keplerian
motion. This means that the data are treated in a homogeneous and comparable way.
Fig.~\ref{fig_mbh} also shows the
black hole mass obtained for Mrk110 via gravitational redshift 
(Kollatschny 2003b). The objects Mrk110 and 3C120 clearly lie above the regression line for quiescent galaxies.

To understand the overly massive black holes of Mrk110 and 3C120, 
we considered three potentially scenarios. 
\begin{itemize} 
\item[i)] If the broad-line profiles in AGN are contaminated by
  turbulence or outflowing winds in the line-emitting region (\citealt{1997ApJ...474...91M}; \citealt{2013A&A...549A.100K}), the emission-line
  velocity dispersion ($\sigma_{V}$) may be increased, leading to an
  overestimate of the black hole mass by means of the virial
  product. However, this does not explain the case of Mrk110. 

\item[ii)] If AGNs have (bipolar) outflows (\citealt{2003A&A...407..461K}; \citealt{2009ApJ...704L..80D}), the host galaxy bulge will get a hole, hence will not
be isotropic. 
If such an AGN/bulge system is seen close to pole-on, the bulge will
appear with a flattened stellar distribution and therefore the stellar
bulge velocity dispersion ($\sigma_{*}$) will be lower (than
that of a quiescent galaxy with a more isotropic bulge). 
However, it remains an open question,
whether $\sigma_{*}$ can be diminished by the factor 2-3 needed to
bring 
3C120 and Mrk110 back on the $M_{BH} - \sigma_{*}$
relation.  

\item[iii)] If we consider two galaxies with the same bulge
  mass ($M_{bulge}$), but with a different black hole mass ($M_{BH}$), the sphere of influence of the galaxy with larger $M_{BH}$ will
  be larger and may lead to stronger or more frequent activity, in
  consequence, leading to an enhanced black hole growth relative to
  the bulge growth. Thus, the assumption that AGNs and quiescent
  galaxies follow the same $M_{BH}-\sigma_{*}$ relationship may not be
  valid in general.

\end{itemize}
For high-redshift quasars a similar deviation from the 
$M_{BH} - \sigma_{*}$ relation has been reported; quasars have a higher $M_{BH}$ by a factor of ten or more 
than quiescent galaxies (\citealt{2010ApJ...714..699W},
their Fig. 8). Furthermore, \citet{2014arXiv1404.1080D} found that BHs in high-redshift radio-loud AGN 
are already, or soon will be, about a factor of ten overly massive compared to their host 
galaxies in terms of expectations from the local $M_{BH} - \sigma_{*}$ relation.
These considerations reveal once more that the widely adopted picture of a co-eval
evolution of AGN and host galaxies needs to be re-examined. Future RM studies of large AGN samples may provide more clues to the 
general BLR geometry, potential outflows, and the $M_{BH} - \sigma_{*}$
relation.

\section{Summary and conclusions}
\label{section_conclusions}

We have investigated two aspects of the efficient method
of photometric reverberation mapping. 

Firstly we
examined how the determination of the mean BLR size is affected when the 
broad emission-line is not fully covered by the chosen narrow-band
filter. By modelling a range of configurations and assuming circular Keplerian orbits, 
we found that for
reasonable line-width band-pass configurations the possible biases for
the mean BLR size are lower than a few percent. Asymmetric cutting of
emission line by the filter has probably even weaker effects. 
In view of the overall measurements errors of typically more than
10\%, any bias in the mean BLR size derived from narrow-band photometric
reverberation mapping appears to be negligible.

  Secondly, we modelled the feature-rich photometric reverberation light
  curves of 3C120 to determine the basic geometry of the BLR, whether it
  is spherical or disk-like. The narrow-band light curves cover the
  central and the blue part of the H$\beta$ line, but miss the red wing.
  Guided by velocity-resolved spectroscopic reverberation data, 
  we assumed that the BLR, as seen in our narrow-band 
  light curves, is dominated by the Keplerian motion and not by in- or
  outflows nor by turbulence. While a spherical BLR model does not fit
  the PRM observations, a thin-disk BLR model that extends from 22 to 28
  light days and has an inclination $i = 10 \pm 4$ degrees
  excellently reproduced the H$\beta$ light curve of 3C120. A
  similar fit quality was reached for a moderately thick-disk BLR model 
  with inclination of $i=5^{\circ}\pm6^{\circ}$, and a covering half-angle 
  $\alpha = 10^{\circ}\pm15^{\circ}$.

 Finally, if this result of a disk-like BLR also holds for Seyfert
 galaxies in general, the determination of the $f$-factor used in
 black hole mass calculations can be remarkably improved. The current
 data already indicate that the black hole of 3C120 is more massive
 than the $M_{BH} - \sigma_{*}$ relation for
 quiescent galaxies. 
 While this needs to be corroborated with future data, 
 it appears to be consistent with findings from the gravitational redshift of
 Mrk110 and with overly massive black holes in high-redshift AGNs.

\begin{acknowledgements}
  
  This work is supported by the
  Nordrhein-Westf\"alische Akademie der Wissenschaften und der K\"unste
  in the framework of the academy program of the Federal Republic of
  Germany and the state Nordrhein-Westfalen and by the Deutsche Forschungsgemeinschaft (DFG HA3555/12-1 and KO857/32-1).  
  The observations on Cerro Armazones benefitted
  from the care of the guardians Hector Labra, Gerardo Pino, Roberto Munoz, 
  and Francisco Arraya. This research has made use of the NASA/IPAC
  Extragalactic Database (NED) which is operated by the Jet Propulsion
  Laboratory, California Institute of Technology, under contract with
  the National Aeronautics and Space Administration. This research has made 
  use of the SIMBAD database, operated at CDS, Strasbourg, France.
  We thank the anonymous referee for constructive comments.

\end{acknowledgements}

\bibliographystyle{aa} 
\bibliography{agn_prm}

\end{document}